\documentclass[preprint]{aastex631}

\begin{document}

\title{The M- and X-class White-light Flares in Super Active Region NOAA 13664/13697 Observed by ASO-S/LST/WST}

\author[0000-0002-8401-9301]{Zhichen Jing}
\author[0000-0002-8258-4892]{Ying Li}
\affiliation{Key Laboratory of Dark Matter and Space Astronomy, Purple Mountain Observatory, CAS, Nanjing 210023, People's Republic of China}
\affiliation{School of Astronomy and Space Science, University of Science and Technology of China, Hefei 230026, People's Republic of China}

\author[0009-0007-7657-1706]{Jingwei Li}
\author[0000-0001-7540-9335]{Qiao Li}
\affiliation{Key Laboratory of Dark Matter and Space Astronomy, Purple Mountain Observatory, CAS, Nanjing 210023, People's Republic of China}

\correspondingauthor{Ying Li}
\email{yingli@pmo.ac.cn}

\begin{abstract}
Solar white-light flares (WLFs) have been observed since 1859, but their occurrence rate is not yet fully understood. The physical properties of WLFs in super active regions (SARs) are also well worth investigating. With full-disk images at 3600 \AA\ (in the Balmer continuum) from the White-light Solar Telescope (WST) on board the Advanced Space-based Solar Observatory, we here study the M- and X-class WLFs occurring in SAR NOAA 13664/13697 (a same region in two solar Carrington rotations) during May/June 2024. 48 WLFs at 3600 \AA\ are identified from 89 available samples with an occurrence rate of 53.9\%, which is much higher than that (23.9\%) of a long-term-continuous data set from October 2022 to May 2023 in our previous work \citep{2024SoPh..299...11J}. In particular, with an additional sample of over 730 M- and X-class flares from October 2022 to June 2024, we find that the occurrence rate of WLFs shows a good correlation with the solar cycle represented by sunspot counts. As regards the properties of the emission at 3600 \AA, the WLFs in SAR NOAA 13664/13697 have some different characteristics, say, a longer duration but a weaker relative enhancement and a smaller brightening area compared with the previous long-term-continuous sample. We also find that for WLFs in NOAA 13664/13697 the relation between the duration and energy at 3600 \AA\ is described by a power-law with index of 0.35, which is similar to the results found for superflares in Sun-like stars \citep{2024LRSP...21....1K}. All these help us understand the solar WLFs in SARs and also provide important insights into the superflares on Sun-like stars.
\end{abstract}


\section{Introduction}
\label{sec:intro}
Solar white-light flares (WLFs) are a kind of flares showing an optical continuum enhancement \citep{1970SoPh...13..471S,1989SoPh..121..261N}, with the first one observed in 1859, i.e., the Carrington event \citep{1859MNRAS..20...13C,1859MNRAS..20...15H}. It has been found that the white-light (WL) emission has a good correlation with the hard X-ray (HXR) emission in many WLFs \citep[e.g.,][]{2006SoPh..234...79H,2016ApJ...816....6K}. These two waveband emissions usually have a similar peak time during the rise phase of the flare in time profile \citep[e.g.,][]{2007ApJ...656.1187F} and in space the WL sources match the HXR footpoint sources generally \citep[e.g.,][]{2011ApJ...739...96K}. Solar WLFs are divided into type I and type II based on observational features \citep[e.g.,][]{1985SoPh...98..255B,1995A&AS..110...99F,2007ASPC..368..417D}. Type I WLFs show a Balmer jump and are usually related to HXR emissions well, while type II WLFs do not have those features. It has been reported that solar WLFs display some similar characteristics with the stellar flares on Sun-like stars, especially in the relationship between flare duration and energy \citep[e.g.,][]{2017ApJ...851...91N,2020ApJ...890...46T,2021MNRAS.505L..79Y}. The power-law index in the duration and energy distribution is 0.39 for some superflares on G-type main sequence stars, which is similar to that of solar flares and close to the theoretical index value of $1/3$ \citep{2015EP&S...67...59M}. Considering that stellar flares are usually observed in the visible passband without spatial resolution, studying solar WLFs in details can thus help understand those superflares.

Only a few hundred WLFs have been reported since the Carrington event \citep[e.g.,][]{1993SoPh..144..169N,2017ApJ...850..204W,2020ApJ...904...96C} and the occurrence rate of WLFs remains a question that needs to be investigated. A few decades ago, \cite{1972SoPh...23..444M} estimated an occurrence rate of WLFs of 5--6 per year near the solar maximum, mainly based on WLF observations at a relatively long wavelength of 5900 \AA. \cite{1983SoPh...88..275N} further derived a higher occurrence rate of $\sim$15 per year for a 2.6-year data set based on observations at $\leq$4000 \AA. In recent years, \cite{2018ApJ...867..159S} reported a high WLF proportion of 36.7\% in circular-ribbon flares using a data set of 90 flares at 6173 \AA\ observed by the Helioseismic and Magnetic Imager \citep[HMI;][]{2012SoPh..275..207S} on board the Solar Dynamics Observatory \citep[SDO;][]{2012SoPh..275....3P}. In addition, \cite{2024ApJ...975...69C} found that the WLF proportion is higher in the confined flares than the eruptive flares across various energy levels based on a few dozens of flares observed by HMI. Thanks to the routine full-disk imaging observations at 3600 \AA\ (Balmer continuum) from the White-light Solar Telescope (WST) of the Ly$\alpha$ Solar Telescope \citep[LST;][]{2019RAA....19..158L,2024SoPh..299..118C} on board the Advanced Space-based Solar Observatory \citep[ASO-S;][]{2023SoPh..298...68G}, the occurrence rate of WLFs in a complete seven-month data set was provided \citep{2024SoPh..299...11J}. 49 WLFs at 3600 \AA\ were identified from 205 M- and X-class flares from October 2022 to May 2023, with an occurrence rate of 23.9\%. It was also found that the WLF occurrence rate increases with the flare magnitude. 

The physical properties of WL emissions have been studied for different WLF samples in recent years. For example, \cite{2018ApJ...867..159S} analyzed 33 circular-ribbon WLFs, which have a mean duration of $\sim$12.5 min and a mean enhancement of 9.4\% for the emission at 6173 \AA. By contrast, the 20 WLFs from active region (AR) NOAA 11515 only have mean duration and enhancement of 4.65 min and 8.1\% at 6173 \AA, respectively \citep{2018A&A...613A..69S}. For the emission at 3600 \AA, \cite{2024SoPh..299...11J} analyzed 39 WLFs during the period of October 2022 and May 2023 and found that they have mean duration, brightening area, and enhancement of 10.3 min, 479 arcsec$^2$, and 19.4\%, respectively. Nevertheless, the properties of WLFs in a special kind of ARs, the so-called super active regions (SARs), and their differences with those of some long-period samples mainly from ordinary ARs are still worthwhile to be investigated. SARs are defined as ARs that can produce more than five flares with the HXR peak count rate exceeding 1000 $\mathrm{count\ s^{-1}}$ in an early time \citep{1987ApJ...314..795B}. Later, \cite{2002SoPh..209..361T} further defined the SAR as an AR with three or more parameters satisfying the following criteria: the largest area $\ge 1000$ millionths of the solar hemisphere (MSH), X-ray flare index $\ge 5.0$, 10.7 cm peak flux $\ge 1000\ \text{solar flux units}\ (\mathrm{s.f.u.})$, proton ($\ge 10\ \mathrm{MeV}$) flux $\ge 400\ \text{proton flux units}\ (\mathrm{p.f.u.})$, and geomagnetic index $A_p \ge 50$. In short, compared with ordinary ARs, SARs are much rarer but produce a large fraction of major flares. Therefore, investigating the properties of WLFs occurring in SARs can provide a deeper insight into the physical mechanisms of WLFs.

SAR NOAA 13664 was visible on the solar disk from 2024 May 2 to 16 and had the largest area of $\sim$2400 MSH, the X-ray index of $\sim$50, and the geomagnetic index $A_p$ of 400. It is one of the largest and most active regions in recent decades \citep[e.g.,][]{2024ApJ...976L..12J, 2024OAP....37..112K, 2025ApJ...979...49H}. This region showed a complex magnetic structure and produced 12 X-class flares, 11 of which are identified as WLFs \citep{2024ApJ...972L...1L}. NOAA 13664 remains to be visible in the next solar Carrington rotation from 2024 May 29 to June 11 and was numbered as NOAA 13697 which also satisfies the criteria of SARs as described above. NOAA 13664 and 13697 produced more than 100 M- and X-class flares in total, which can serve as a good flare data set for studying the properties of SARs.

In this paper, we analyze the M- and X-class flares in SAR NOAA 13664/13697 observed by ASO-S/LST/WST and study the occurrence rate of WLFs at 3600 \AA. The properties of the 3600 \AA\ emission including relative enhancement, brightening area, duration, and energy are also investigated and compared with those of the WLFs during October 2022 and May 2023 (referred to as the long-term-continuous sample hereafter) as reported in \cite{2024SoPh..299...11J}. All these can greatly help us understand the occurrence rate and observational properties of WLFs in SARs and further assist in the investigation of superflares on Sun-like stars.

\section{Observational Data and Flare Data Set} \label{sec:data}

The WL data used in this work are from WST on ASO-S. WST provides continuous full-disk images at 3600$\pm20$ \AA\ with a pixel size of $\sim$0.5$^{\prime\prime}$ but a spatial resolution of $\sim$4$^{\prime\prime}$. The time cadence of WST routine observations is 2 min. Here we analyze about 5000 images for $\sim$100 M- and X-class flares from NOAA 13664/13697 and additional $\sim$35000 images for another 600+ flares from WST. All these images have been corrected for dark current and flat field as well as made registration including north up and translation \citep{2024SoPh..299..118C}. The Hard X-ray Imager \citep[HXI;][]{2019RAA....19..160Z} also on ASO-S has an energy band of $\sim$10--300 keV and a temporal resolution better than 0.5 s. Its HXR 20--50 keV data for several tens of WLFs are used here. The HXR light curves have been produced with energy calibration and preprocessing including the correction of differential nonlinearity effect and detector response matrix \citep{2024SoPh..299..153S}. We also use the soft X-Ray (SXR) 1--8 \AA\ data observed by the Geostationary Operational Environmental Satellite \citep[GOES;][]{1996SPIE.2812..344H}, which are widely used to define the flare class. 

There were 90 M-class and 17 X-class flares recorded by GOES from NOAA 13664 and 13697. Among them, 3 M-class flares were completely missed by WST due to the Earth eclipse and 15 M-class flares were observed by WST only covering a partial time without any brightening detected using the 2-min cadence data. These 18 flares are excluded from our data set. As a result, we have 89 (72 M-class and 17 X-class) flares to study the occurrence rate of WLFs (in Section \ref{subsub:rate}), in which 66 (54 M-class and 12 X-class) are from NOAA 13664 and 23 (18 M-class and 5 X-class) are from NOAA 13697 (listed in Table \ref{tab:1}). Note that in the data set there include some flares only with partially available WST data but showing excess emissions at 3600 \AA. When identifying a WLF at 3600 \AA, we use the same method as that in our previous work \citep{2024SoPh..299...11J}, i.e., an intensity ratio defined as $(I-I_0)/I_{\rm background}$, where $I$, $I_0$, and $I_{\rm background}$ represent the intensity of a pixel in the flare time, the mean intensity of the same pixel over 30 min before the flare onset, and the mean intensity of a quiet-Sun region before the flare onset, respectively. The threshold of the intensity ratio is adopted the same as 8\% for the 3600 \AA\ emission.

From the 48 identified WLFs in NOAA 13664 and 13697 (see Table \ref{tab:1}), we further select 22 (14 M- and 8 X-class, listed in Table \ref{tab:2}) that have available WST data covering the entire flare time to study the physical parameters of 3600 \AA\ emission (in Section \ref{subsec:parameter}) including brightening area $S$, duration $\tau$, maximum pixel enhancement $r_p$, maximum mean enhancement $r_m$, and energy $E$. Note that 10 limb WLFs with longitude larger than 85$^\circ$ have been excluded due to a possible underestimation of the WL area. The definitions of these parameters are the same as those in \cite{2024SoPh..299...11J}. In brief, $S$ stands for the area of all the brightening pixels. $\tau$ represents the interval between the WL start and end times. The relative enhancement $r$ is defined as $(I-I_0)/I_0$, and $r_p$ and $r_m$ represent the maximum enhancement of the brightening pixels and the maximum value of the mean enhancement of the WL brightening region during the flare, respectively. $E$ is an integration of the excess intensity (with radiometric calibration) over area and time. These parameters are measured with the limb darkening and projection effect removed. More details can be found in \cite{2024SoPh..299...11J}.

In this study, we compare the results of the WLFs in NOAA 13664 and 13697 with those of the long-term-continuous sample in \cite{2024SoPh..299...11J}. For the long-term-continuous sample, it contains 205 M- and X-class flares occurring during October 2022 and May 2023. There are 49 out of them (23.9\%) identified as WLFs at 3600 \AA. Among the 49 WLFs, 39 are further selected to study the 3600 \AA\ emission parameters including $r_{p}$, $r_{m}$, $S$, $\tau$, and $E$.

\section{Results} 
\label{sec:results}

\subsection{Overview of the Flare Events in NOAA 13664 and 13697}

\begin{figure*}[ht!]
\centering
\includegraphics[width=\linewidth]{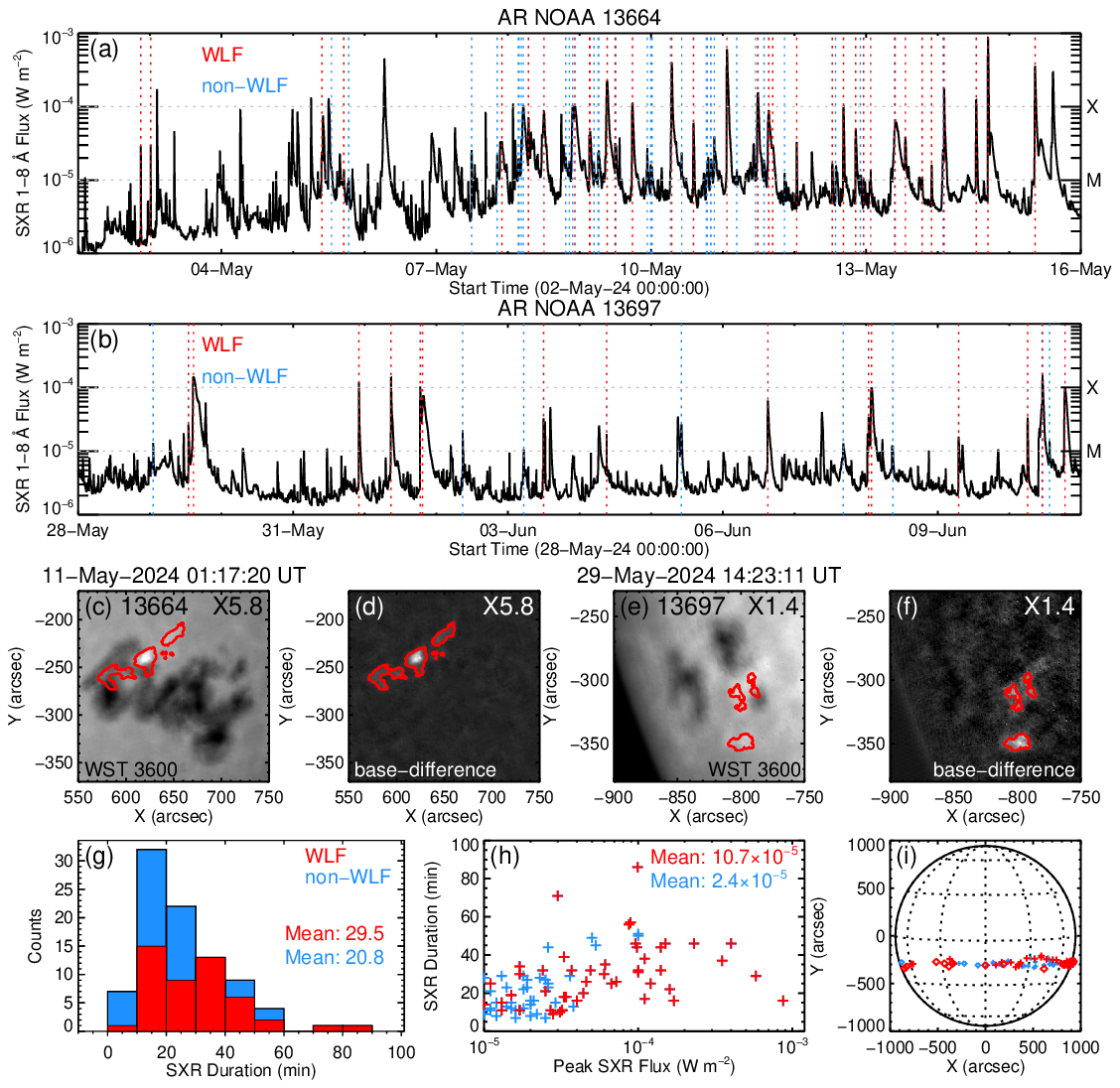}
\caption{Overview of the flare events in NOAA 13664 and 13697. (a) and (b) GOES SXR 1--8 \AA\ flux during the appearing time of NOAA 13664 and 13697. The red and blue vertical lines indicate the WLFs at 3600 \AA\ and non-WLFs from the two regions, respectively. (c)--(f) WST 3600 \AA\ images and its base-difference ones for two example WLFs from the two regions. The WL brightenings are outlined by red contours, which are derived from the base-difference images. (g) Histogram of SXR duration with red and blue colors representing the WLFs and non-WLFs, respectively. (h) Scatter plot of SXR duration versus peak SXR flux with red and blue plus symbols indicating the WLFs and non-WLFs, respectively. (i) Spatial distribution of WLFs (red color) and non-WLFs (blue color). The plus and diamond symbols denote the flares from NOAA 13664 and 13697, respectively.
\label{fig:1}}
\end{figure*}

Figures \ref{fig:1}(a) and (b) show the GOES SXR 1--8 \AA\ flux with the M- and X-class flares occurring in NOAA 13664 and 13697 indicated by red (for identified WLFs at 3600 \AA) and blue (for non-WLFs) vertical lines. One can see that NOAA 13664 produced more numerous and energetic flares than NOAA 13697. Figures \ref{fig:1}(c)--(f) present the WST 3600 \AA\ plus its base-difference images for two example WLFs from the two regions. We can see some brightenings (marked by red contours) appear near or in the sunspots. It is obvious that NOAA 13664 (Figure \ref{fig:1}(c)) has a more complex structure composed of multiple sunspots with complicated shapes and has a larger sunspot area than NOAA 13697 (Figure \ref{fig:1}(e)).

We also compare the SXR duration and the peak SXR flux between WLFs and non-WLFs in NOAA 13664 and 13697. From Figure \ref{fig:1}(g) one can see that the WLFs in these two regions have a mean SXR duration of 29.5 min, which is longer than that (20.8 min) of non-WLFs. Note that this result is contrary to the case of the long-term-continuous sample in \cite{2024SoPh..299...11J} where WLFs have a shorter mean SXR duration compared with non-WLFs. For the peak SXR flux, the WLFs from the two regions have a larger value than non-WLFs, as shown in Figure \ref{fig:1}(h). This is consistent with the result in the long-term-continuous sample and also reflects that a larger flare is more likely to be a WLF as reported before. Figure \ref{fig:1}(i) shows the spatial distribution of all WLFs and non-WLFs in NOAA 13664 (plus symbols) and 13697 (diamond symbols). It is seen that there shows an asymmetry in the distribution of flares for the western and eastern hemispheres \citep{2022Ge&Ae..62..288P} and that WLFs tend to occur near the solar limb rather than in the disk center \citep{2024SoPh..299...11J}.

\subsection{Occurrence Rate of WLFs at 3600 \AA}
\label{subsec:occurrence}

\subsubsection{Occurrence Rate of WLFs in NOAA 13664 and 13697}
\label{subsub:rate}

As demonstrated in Table \ref{tab:1}, 33 out of 66 flares are identified as WLFs for NOAA 13664, with an occurrence rate of 50.0\%. While a higher occurrence rate (65.2\%) is found in NOAA 13697, where 15 WLFs are identified from 23 flares. The occurrence rate increases with flare magnitude for both regions, which is consistent with the result of the long-term-continuous sample in \cite{2024SoPh..299...11J}. In NOAA 13664, the occurrence rate of WLFs for M- and X-class flares are 42.6\% and 83.3\%, respectively, which are 55.6\% and 100\% respectively in NOAA 13697. When considering the two regions as a whole group, the occurrence rate of WLFs for M- and X-class flares are 45.8\% and 88.2\%, respectively, and the total occurrence rate is 53.9\% with 48 WLFs identified from 89 flares in total. The total occurrence rate in NOAA 13664/13697 is much higher than that (23.9\%) in the long-term-continuous sample. The main reason is supposed to be a high proportion (19.1\%) of the X-class flares in NOAA 13664 and 13697, which is much higher than that (only 3.4\%) in the long-term-continuous sample. 

\begin{table*}[h]
\caption{Occurrence rate of WLFs in NOAA 13664 and 13697}
\raggedright
\label{tab:1}
\setlength{\tabcolsep}{1.0mm}{
\begin{tabular}{c|ccc|ccc|ccc}     
\hline 
\hline
 AR & & NOAA 13664 & & & NOAA 13697 & & & Total &  \\
\cline{2-10} 
  & M-class & X-class & Total & M-class & X-class & Total & M-class & X-class & Total \\
\hline
WLF & 23 & 10 & 33 & 10 & 5 & 15 & 33 & 15 & 48\\
non-WLF & 31 & 2 & 33 & 8 & 0 & 8 & 39 & 2 & 41\\
Total & 54 & 12 & 66 & 18 & 5 & 23 & 72 & 17 & 89\\
\hline
Occurrence Rate & 42.6\% & 83.3\% & 50.0\% & 55.6\% & 100\% & 65.2\% & 45.8\% & 88.2\% & 53.9\%\\
\hline
\hline
\end{tabular}
}
\end{table*}

\subsubsection{Occurrence Rate of WLFs during the Period of October 2022 and June 2024}
\label{subsub:rate long}

Considering that NOAA 13664 and 13697 appear in the rise phase of solar cycle 25, the relatively high occurrence rate of WLFs may be related to the whole level of solar activity that can be represented by sunspot counts. Here we investigate the relationship between the WLF occurrence rate and sunspot number\footnote{The data of sunspot counts come from the website of \href{https://www.swpc.noaa.gov/products/solar-cycle-progression}{Space Weather Prediction Center of NOAA}.} for the period from October 2022 to June 2024, namely from the first light of WST to the appearing time of two ARs under study. During this 20-month period, there were $\sim$800 M- and X-class flares recorded by GOES and 738 of them were well observed by WST in a routine mode. From these 738 flares, we identify 294 WLFs at 3600 \AA, with a total occurrence rate of 39.8\%, which is actually higher than that (23.9\%) for the long-term-continuous sample \citep{2024SoPh..299...11J}. Figures \ref{fig:2}(a) and (b) show the sunspot number and WLF occurrence rate, respectively, over the 20-month period with an interval of one month, together with the time profile of the smoothed sunspot number. One can see that the sunspot number exhibits a growing trend with some fluctuations and the WLF occurrence rate displays a similar increasing but kind of delayed trend from 16.7\% to 59.0\% during the period. We also show the scatter plot between the WLF occurrence rate and the smoothed sunspot number in Figure \ref{fig:2}(c). It is seen that these two parameters have a good relationship with the Pearson correlation coefficient (Pcc) of 0.64 and a significance of 0.0025. This suggests that the occurrence rate of WLFs is generally becoming higher and higher as the solar maximum approaches. It should be mentioned that if we delay several months of the WLF occurrence rate, its correlation with sunspot number can be Pcc$>$0.80 with a significance of 10$^{-5}$. In Figures \ref{fig:2}(b) and (c), we over-plot the WLF occurrence rates of NOAA 13664 and 13697 (see the red and blue asterisks). These rates are even higher than those of May and June 2024, respectively. This implies that SARs may be more likely to produce WLFs than ordinary ARs.

\subsection{Physical Properties of WLFs in NOAA 13664 and 13697} 
\label{subsec:parameter}

Besides the WLF occurrence rate, we study the properties of 3600 \AA\ emission for the WLFs in SAR NOAA 13664/13697. Here we select 22 WLFs from the 48 identified WLFs, which have full WST data for the entire flare time, to measure the maximum pixel enhancement ($r_{p}$), maximum mean enhancement ($r_{m}$), brightening area ($S$), duration ($\tau$), and energy ($E$) at 3600 \AA. The results are listed in Table \ref{tab:2}. Figure \ref{fig:3} exhibits the histograms of $r_{p}$, $r_{m}$, $S$, and $\tau$ for the WLFs from NOAA 13664/13697 (red color), together with those of the long-term-continuous sample in \cite{2024SoPh..299...11J} (green color) for comparison. We can see that most WLFs (more than two thirds) in NOAA 13664/13697 have values of $r_{p}$ and $r_{m}$ less than 30\% and 15\%, respectively (Figures \ref{fig:3}(a) and (b)), while the long-term-continuous sample has greater values mostly. The mean (median) values of $r_p$ and $r_m$ of the WLFs in NOAA 13664/13697 are 37.4\% (25.0\%) and 14.0\% (12.2\%), respectively, i.e., smaller than those, namely 41.7\% (32.3\%) and 19.4\% (17.4\%), of the long-term-continuous sample. Similarly, the mean and median values of $S$ (391 and 138 arcsec$^{2}$ respectively) for the WLFs in NOAA 13664/13697 are smaller than those (479 and 225 arcsec$^{2}$ respectively) of the long-term-continuous sample (Figure \ref{fig:3}(c)). However, there shows a different result in $\tau$. From Figure \ref{fig:3}(d) we can see that the mean and median values of $\tau$ for the WLFs in NOAA 13664/13697 are 12.9 and 8.2 min, respectively, which are longer than those (10.3 and 7.8 min) of the long-term-continuous sample. In a word, the 22 WLFs in SAR NOAA 13664/13697 have smaller $r_{p}$, $r_{m}$, and $S$ but longer $\tau$ than those of the long-term-continuous sample.

\begin{table*}[htb]
\caption{Information of the 22 WLFs analyzed in detail}
\label{tab:2}
\tiny
\setlength{\tabcolsep}{1.0mm}{
\begin{tabular}{cccccccccccccccc}                               
\hline                   
 AR & Observation & Flare & GOES & GOES & HXR & WL & WL & WL & WL & WL & WL & WL & WL & WLF\\
 NOAA & Date & Location & Class & Peak & Peak\footnote{The HXR peak time is represented by the peak time of the HXR emission at 20--50 keV. Note that the HXR peak time of the M3.3 flare on 2024 May 7 is tagged as ``-" since the HXR emission observed by HXI is contaminated by South Atlantic Anomaly.} & Start & Peak & End & $\tau$ & $r_\mathrm{p}$ & $r_\mathrm{m}$\footnote{The uncertainty of $r_{m}$ is estimated to be  1.39\%, which is the mean value of the quiet-Sun intensity fluctuation.} & $S$ & $E$\footnote{The uncertainty of $E$ comes from the radiometric calibration.} & Group\\
 & (yyyy-mm-dd) &  &  & (UT) & (UT) & (UT) & (UT) & (UT) & (min) & (\%) & (\%) & (arcsec$^2$) & (erg) & \\
\hline
13664 & 2024-05-02 & S20E59 & M2.8 & 20:57 & 20:55 & 20:55 & 20:55 & 20:59 & 4.7 & 21.3 & 12.0 & 132 & (1.30$\pm$0.18)$\times10^{28}$ & 1\\
 & 2024-05-03 & S20E57 & M2.7 & 00:15 & 00:14 & 00:16 & 00:16 & 00:18 & 2.0 & 14.2 & 10.6 & 11.0 & (1.31$\pm$0.14)$\times10^{27}$ & 2\\
 & 2024-05-05 & S21E22 & M2.4 & 09:38 & 09:28 & 09:35 & 09:37 & 09:39 & 4.0 & 22.9 & 12.4 & 10.8 & (7.67$\pm$1.10)$\times10^{26}$ & 2\\
 & 2024-05-07 & S22W07 & M3.3 & 21:59 & - & 21:48 & 21:52 & 22:04 & 16 & 27.1 & 10.5 & 92.0 & (1.18$\pm$0.17)$\times10^{28}$ & 1\\
 & 2024-05-08 & S20W17 & M7.2 & 06:53 & 06:49 & 06:50 & 06:50 & 06:52 & 2.0 & 18.8 & 10.6 & 12.3 & (4.00$\pm$0.59)$\times10^{26}$ & 1\\
 & 2024-05-08 & S20W11 & M8.7 & 12:04 & 12:00 & 11:37 & 12:03 & 12:23 & 46 & 36.7 & 12.3 & 203 & (5.27$\pm$0.76)$\times10^{28}$ & 2\\
 & 2024-05-09 & S21W23 & M4.0 & 03:17 & 03:15 & 03:15 & 03:17 & 03:19 & 4.0 & 27.5 & 19.5 & 18.2 & (1.54$\pm$0.22)$\times10^{27}$ & 1\\
 & 2024-05-09 & S22W23 & M4.6 & 03:32 & 03:29 & 03:33 & 03:39 & 03:51 & 18 & 22.8 & 12.1 & 60.6 & (1.06$\pm$0.15)$\times10^{28}$ & 2\\
 & 2024-05-09 & S20W24 & X2.3 & 09:13 & 09:04 & 09:02 & 09:22 & 09:41 & 39 & 44.7 & 13.0 & 783 & (3.76$\pm$0.41)$\times10^{29}$ & 2\\
 & 2024-05-09 & S16W39 & M3.1 & 11:56 & 11:54 & 11:54 & 11:54 & 11:55 & 1.2 & 18.9 & 10.5 & 203 & (6.68$\pm$0.72)$\times10^{27}$ & 1\\
 & 2024-05-09 & S17W28 & X1.1 & 17:44 & 17:32 & 17:30 & 17:32 & 17:40 & 9.8 & 27.4 & 11.6 & 558 & (5.18$\pm$0.56)$\times10^{28}$ & 1\\
 & 2024-05-10 & S17W34 & X4.0 & 06:54 & 06:43 & 06:35 & 06:43 & 07:04 & 29 & 120 & 12.6 & 1630 & (4.39$\pm$0.47)$\times10^{29}$ & 1\\
 & 2024-05-11 & S17W44 & X5.8 & 01:23 & 01:17 & 01:17 & 01:17 & 01:35 & 18 & 94.1 & 16.8 & 2224 & (9.21$\pm$0.99)$\times10^{29}$ & 1\\
 & 2024-05-12 & S20W75 & X1.0 & 16:26 & 16:19 & 16:20 & 16:20 & 16:42 & 22 & 61.9 & 22.5 & 568 & (1.93$\pm$0.25)$\times10^{29}$ & 1\\
 & 2024-05-12 & S20W78 & M4.9 & 20:32 & 20:25 & 20:25 & 20:25 & 20:27 & 2.0 & 82.5 & 28.3 & 305 & (4.35$\pm$0.56)$\times10^{28}$ & 1\\
 & 2024-05-12 & S20W79 & M1.0 & 23:10 & 23:09 & 23:00 & 23:12 & 23:20 & 20 & 14.6 & 10.5 & 44.7 & (3.48$\pm$0.60)$\times10^{28}$ & 2\\
\hline
13697 & 2024-05-29 & S20E66 & X1.4 & 14:37 & 14:23 & 14:23 & 14:23 & 14:35 & 12 & 22.2 & 9.5 & 1289 & (2.37$\pm$0.42)$\times10^{29}$ & 1\\
 & 2024-05-31 & S17E34 & X1.1 & 22:03 & 22:01 & 21:58 & 22:04 & 22:12 & 14 & 65.2 & 20.9 & 144 & (2.90$\pm$0.51)$\times10^{28}$ & 2\\
 & 2024-06-01 & S18E22 & X1.0 & 18:36 & 18:32 & 18:35 & 18:35 & 18:41 & 6.5 & 29.4 & 16.7 & 90.7 & (7.31$\pm$1.29)$\times10^{27}$ & 2\\
 & 2024-06-03 & S19W00 & M3.2 & 11:55 & 11:52 & 11:55 & 11:55 & 11:57 & 2.0 & 18.2 & 11.4 & 45.4 & (2.97$\pm$0.48)$\times10^{27}$ & 2\\
 & 2024-06-04 & S18W12 & M1.7 & 09:04 & 09:02 & 09:01 & 09:02 & 09:06 & 4.9 & 16.5 & 12.8 & 9.0 & (1.40$\pm$0.27)$\times10^{27}$ & 1\\
 & 2024-06-08 & S18W63 & M3.4 & 00:51 & 00:50 & 00:49 & 00:50 & 00:54 & 5.2 & 16.8 & 10.0 & 170 & (1.00$\pm$0.19)$\times10^{28}$ & 1\\
  \hline
\end{tabular}
}
\end{table*}

We further study how the parameters of 3600 \AA\ emission vary with flare magnitude. Figures \ref{fig:4}(a)--(e) exhibit the scatter plots of $r_{p}$, $r_{m}$, $S$, $\tau$, and $E$ versus peak SXR flux for the 22 WLFs in NOAA 13664 and 13697. It is seen that all the parameters except $r_{m}$ have a moderate to strong positive correlation with the peak SXR flux (Pcc$\ge$0.42 with a significance of less than 0.05, see the values in red in the panels), namely $r_{p}$, $S$, $\tau$, and $E$ increase with flare magnitude. It should be mentioned that for the long-term-continuous sample in \cite{2024SoPh..299...11J}, all the parameters including $r_{m}$ have a good correlation with the peak SXR flux with Pcc$\ge$0.58 and a significance of less than 10$^{-4}$ (see the values in green for reference). In addition, we show the scatter plots of $E$ with $r_{p}$, $r_{m}$, $S$, and $\tau$ in Figures \ref{fig:4}(f)--(j). One can see that $E$ has some correlations with $r_{p}$ (Pcc=0.69 with a significance of 0.0004), $S$ (Pcc=0.94 with a significance of 10$^{-10}$), and $\tau$ (Pcc=0.42 with a significance of 0.05) but not with $r_{m}$ (Pcc=0.12) in a linear scale, which is a little different from the result in the long-term-continuous sample (indicated by the Pcc values in green) as well. In particular, $E$ has a better correlation with $\tau$ (Pcc=0.72 with a significance of 0.0002) in the log-log scale as seen in Figure \ref{fig:4}(j). More interestingly, the fitted power-law index is 0.35, i.e., $\tau \propto E^{0.35}$, which is very close to the theoretical value of 1/3 as well as similar to that ($\sim$0.39) for stellar flares \citep[e.g.,][]{2015EP&S...67...59M,2024LRSP...21....1K}. This result, however, is not found in the long-term-continuous sample. 

\begin{figure*}[ht!]
\centering
\includegraphics[width=\linewidth]{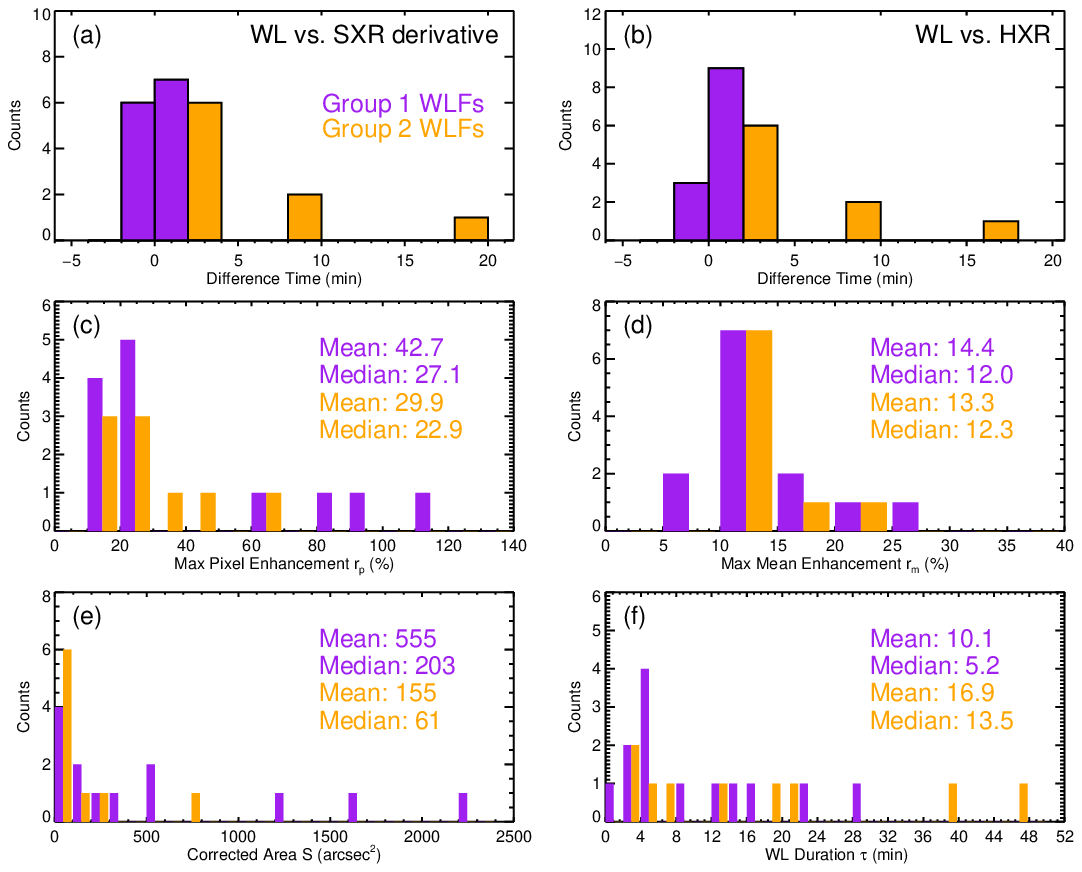}
\caption{(a) and (b) Histograms of the peak-time difference between WL and the SXR time derivative or HXR. Purple and orange colors represent Group 1 and 2 WLFs, respectively. (c)--(f) Distributions of the maximum pixel enhancement ($r_{p}$), maximum mean enhancement ($r_{m}$), corrected area ($S$), and WL duration ($\tau$) of Group 1 (purple) and 2 (orange) WLFs. 
\label{fig:5}}
\end{figure*}

Finally, we investigate relationships between the peak times of the WL at 3600 \AA\ and SXR 1--8 \AA/HXR 20--50 keV emissions for the 22 WLFs in NOAA 13664 and 13697. Figure \ref{fig:5}(a) shows the histogram of the peak-time difference between WL and the SXR time derivative. One can see that more than half (13) WLFs have a similar peak time (within $\pm$2 min according to the cadence of WST routine mode) between the two emissions, while the remaining (9) WLFs have a WL peak time later than that of SXR derivative. Here we refer to the former 13 WLFs as Group 1, whose WL emissions hold the Neupert effect \citep{1968ApJ...153L..59N}, and the remaining 9 WLFs as Group 2 (also indicated in Table \ref{tab:2}). As expected, these two groups are well displayed in the histogram of the peak-time difference between WL and HXR (see Figure \ref{fig:5}(b)), namely Group 1 has a similar peak time between WL and HXR as well, while Group 2 exhibits a later peak time in WL than in HXR. Based on the relationship between the WL and HXR peak times, we could say that Group 1 WLFs are likely to be the traditional type I WLFs, but further checking of the Balmer jump is still needed. Figures \ref{fig:5}(c)--(f) plot the histograms of $r_p$, $r_m$, $S$, and $\tau$ for Group 1 (purple) and 2 (orange) WLFs for comparison. We can find that Group 1 WLFs have larger $r_p$, $r_m$, and $S$, but shorter $\tau$ than Group 2 WLFs overall (see the mean and median values). This result is actually the same as that for the long-term-continuous sample in \cite{2024SoPh..299...11J}.  We also find that the proportion of Group 1 WLFs is only 59.1\% (13 out of 22) in SAR NOAA 13664/13697, which is much lower than that (82.1\%, 32 out of 39) in the long-term-continuous sample. This can to some extent explain why the WLFs in NOAA 13664 and 13697 have smaller $r_p$, $r_m$, and $S$ but longer $\tau$ as described above.

\section{Summary and Discussions} \label{sec:summary}

In this work, we analyze a few tens of M- and X-class flares occurring in SAR NOAA 13664/13697 during May/June 2024 observed by ASO-S/LST/WST at 3600 \AA. To our best knowledge, this is the first time to provide the occurrence rate of WLFs and physical properties of the WLFs from SAR, at least in the Balmer continuum. We also compare the results from SAR with those of a seven-month long-term-continuous sample in \cite{2024SoPh..299...11J}. The main results are summarized below.

\begin{figure*}[ht!]
\centering
\includegraphics[width=\linewidth]{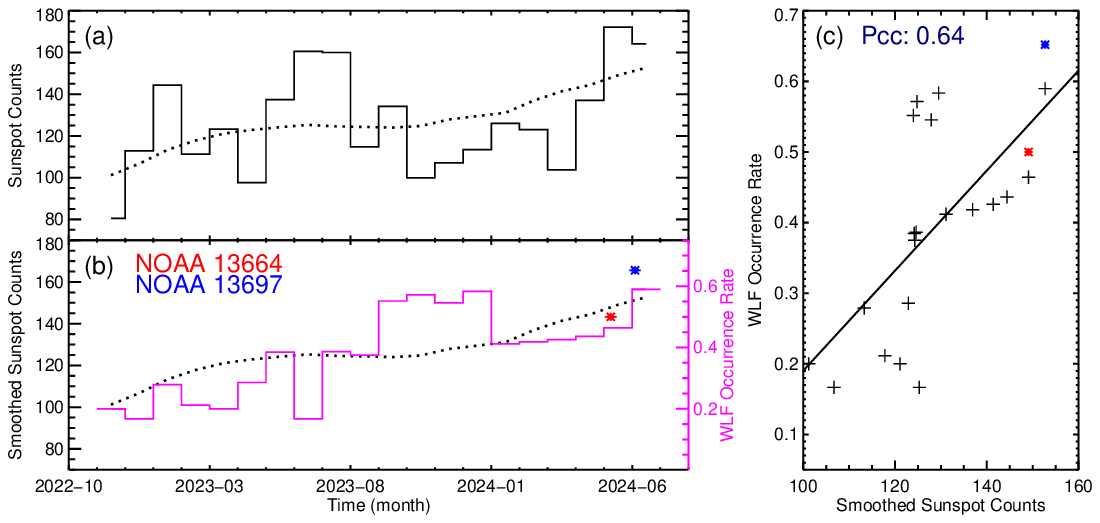}
\caption{(a) Time profile of the sunspot number (solid curve) and its smoothed profile (dotted curve) during the period of October 2022 and June 2024. (b) Temporal evolution of the WLF occurrence rate (magenta curve, corresponding to the right axis), together with the smoothed sunspot number (dotted curve) during the period. The red and blue asterisks in this panel (and also in panel (c)) mark the occurrence rates of WLFs in NOAA 13664 and 13697, respectively. (c) Scatter plot of the WLF occurrence rate versus the smoothed sunspot number. The straight line is their linear fit. 
\label{fig:2}}
\end{figure*}

\begin{figure*}[ht!]
\centering
\includegraphics[width=\linewidth]{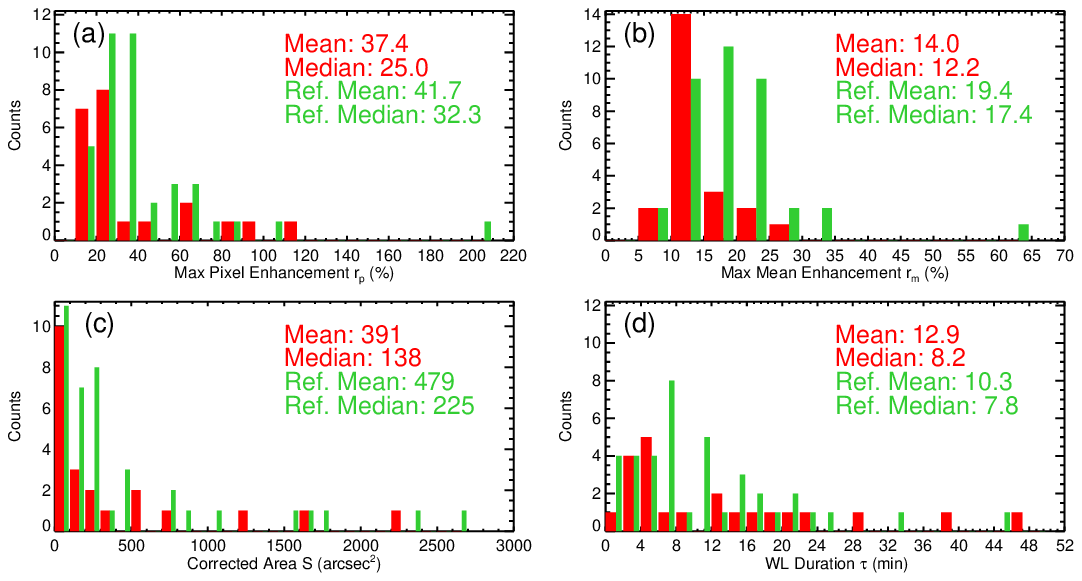}
\caption{Histograms of the maximum pixel enhancement ($r_{p}$) (a), maximum mean enhancement ($r_{m}$) (b), brightening area ($S$) corrected for projection effect (c), and WL duration ($\tau$) at 3600 \AA\ (d). Results for the 22 WLFs from NOAA 13664 and 13697 are shown in red, and those for the long-term-continuous sample in \cite{2024SoPh..299...11J} are in green for comparison. 
}
\label{fig:3}
\end{figure*}

\begin{figure*}[ht!]
\centering
\includegraphics[width=\linewidth]{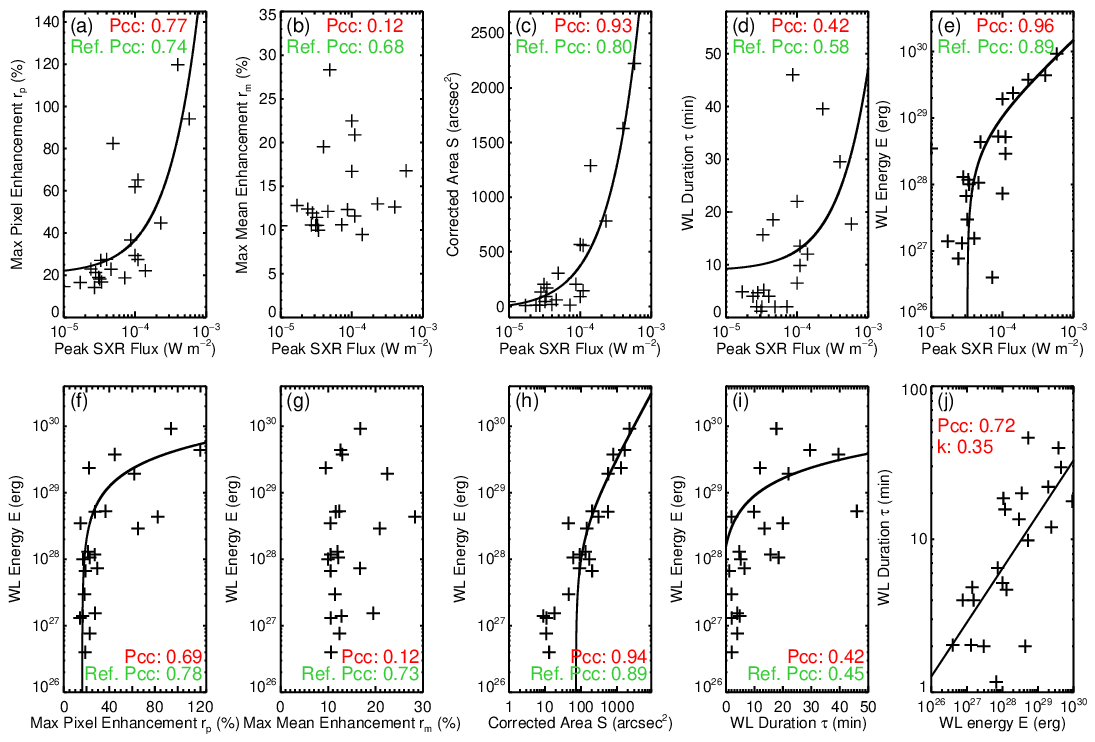}
\caption{Top panels show the relationships of maximum pixel enhancement ($r_{p}$), maximum mean enhancement ($r_{m}$), corrected area ($S$), WL duration ($\tau$), and WL energy ($E$) with peak SXR flux for the 22 WLFs in NOAA 13664 and 13697. Bottom panels exhibit the relationships of $E$ with $r_{p}$, $r_{m}$, $S$, and $\tau$. When the Pcc value (in red) is greater than 0.40, we make a linear fit (see the black curve). The Pcc values in green denote the results of the long-term-continuous sample in \cite{2024SoPh..299...11J} for reference.
}
\label{fig:4}
\end{figure*}

\begin{enumerate}
\item There are 48 WLFs identified from 89 M- and X-class flares in SAR NOAA 13664/13697, with an occurrence rate of 53.9\% (Table \ref{tab:1}, Section \ref{subsub:rate}), which is much higher than that (23.9\%, 49 out of 205) of the long-term-continuous sample. This high occurrence rate is related to a high proportion of the X-class flares in NOAA 13664 and 13697, as the WLF occurrence rate increases with the flare magnitude \citep[e.g.,][]{2024SoPh..299...11J}, which is also the case for each of the two regions.
\item The WLF occurrence rate becomes higher and higher in general as the solar maximum approaches, as indicated by a good correlation between the occurrence rate and sunspot number during a 20-month period (Figure \ref{fig:2}, Section \ref{subsub:rate long}). In particular, the occurrence rates of WLFs in NOAA 13664 and 13697 are even higher than the monthly rates of May and June 2024, respectively, implying that SARs may be more likely to produce WLFs than ordinary ARs.
\item The 22 selected WLFs in NOAA 13664 and 13697 have smaller relative enhancement and brightening area but longer duration at 3600 \AA\ than the 39 selected WLFs in the long-term-continuous sample (Figure \ref{fig:3}, Section \ref{subsec:parameter}). This can be explained to a certain extent by a low proportion of Group 1 WLFs that have around the same peak time in WL and HXR and show relatively larger enhancement and area but shorter duration. 
\item The 3600 \AA\ emission parameters including maximum pixel enhancement, brightening area, duration, and energy increase with flare magnitude for the WLFs in NOAA 13664 and 13697 (Figures \ref{fig:4}(a)--(e), Section \ref{subsec:parameter}). The energy at 3600 \AA\ also has a positive correlation with the other parameters (Figures \ref{fig:4}(f)--(i)). In particular, the energy and duration have a good power-law relationship with an index of 0.35 ($\tau \propto E^{0.35}$) (Figure \ref{fig:4}(j)), which is very close to the theoretical value of 1/3 as well as similar to that for superflares on Sun-like stars.
\end{enumerate}

In addition to a total WLF occurrence rate of 53.9\%, NOAA 13664 and 13697 have occurrence rates of 50.0\% and 65.2\%, respectively, as mentioned in Section \ref{subsub:rate}. The higher occurrence rate of NOAA 13697 may be associated with a higher proportion of confined flares, in which WLFs are more common than in eruptive flares \citep{2024ApJ...975...69C}. The proportions of confined flares are found to be 53.0\% (35 out of 66) in NOAA 13664 and 69.6\% (16 out of 23) in NOAA 13697, according to the coronal mass ejection catalog\footnote{\url{https://cdaw.gsfc.nasa.gov/CME_list/}. This CME catalog is generated and maintained at the CDAW Data Center by NASA and The Catholic University of America in cooperation with the Naval Research Laboratory.} from the Large Angle and Spectrometric Coronagraph (LASCO; \citealt{1995SoPh..162..357B}) onboard the Solar and Heliospheric Observatory (SOHO). The confined flares seem to occur more frequently in the decline phase of ARs \citep{2024ApJ...976L...2L}, which is also the case here. NOAA 13697 appeared in the second Carrington rotation and had a less complex magnetic configuration and a smaller area of sunspots\footnote{As inferred from visual inspection of HMI observations, publicly available at https://solarmonitor.org.} compared with NOAA 13664. We could say that NOAA 13664 and 13697 correspond to the developing and decaying phases of the same AR, respectively, with the latter one showing a higher occurrence rate of WLFs. Note that we cannot exclude the sample bias due to only 23 flares from NOAA 13697 in the role of its higher WLF occurrence rate. 

The WLFs in SAR NOAA 13664/13697 exhibit some different characteristics from the long-term-continuous sample, such as a higher occurrence rate and a longer WL duration, but a lower relative enhancement and a smaller brightening area. In addition, the WL duration and energy for WLFs in SAR NOAA 13664/13697 show a good power-law relationship ($\tau \propto E^{0.35}$), which does not exist for WLFs in the long-term-continuous sample. Note that this power-law relationship has also been derived from the superflares on Sun-like stars \citep[e.g.,][]{2015EP&S...67...59M, 2024LRSP...21....1K}. Considering that the WLFs in the long-term-continuous sample are mainly from ordinary ARs, we may say that the SAR 13664/13697 has some unique properties compared with ordinary ARs, which, however, are more similar to those of ARs on the Sun-like stars. It should be mentioned that we here only investigate a single solar SAR that has been observed by WST. As the solar maximum is approaching, more SARs are needed to study when they show up in the coming years.

In this work, we investigate the WLFs in SAR 13664/13697 only at 3600 \AA, i.e., in the Balmer continuum, while it is important to study these WLFs combining with the Paschen continuum (above 3646 \AA). In fact, Cai et al. (in preparation) are analyzing all the M- and X-class WLFs from this SAR by using the HMI data at 6173 \AA. They use a different method \citep{2024ApJ...975...69C} to identify the WLFs as well as to explore the dynamics of identified WL kernels. A comparison of the WL properties in the Balmer and Paschen continua is worthy of study in the future.

Another interesting thing is the relationship between the WLF occurrence rate and sunspot counts or solar cycle. In the present work, we explore this relationship during a complete 20-month period from October 2022 to June 2024 using the WST data at 3600 \AA. From July 2024 to the current time of writing, there are more than 700 flares above M1.0 recorded by GOES. This extended data set is worthwhile to be considered in the future to study the WLF occurrence rate over the solar cycle.

\begin{acknowledgments}
We sincerely thank the referee for the constructive and insightful suggestions and comments, which improved the quality of our manuscript. The ASO-S mission is supported by the Strategic Priority Research Program on Space Science, Chinese Academy of Sciences. We thank Dr. Yijun Hou, Dechao Song, Yingjie Cai, Jun Tian, Zhengyuan Tian, and Wenhui Yu very much for their helpful discussions. This work is supported by NSFC under grants 12273115 and 12233012, by the Strategic Priority Research Program of the Chinese Academy of Sciences under grant XDB0560000, and by the National Key R\&D Program of China under grants 2022YFF0503004.
\end{acknowledgments}

\newpage

\vspace{5mm}

\bibliography{refer}{}
\bibliographystyle{aasjournal}

\end{document}